# Broadband frequency comb generation in aluminum nitride-on-sapphire microresonators

Xianwen Liu,[1] Changzheng Sun,[1, *] Bing Xiong,[1] Lai Wang,[1] Jian Wang,[1] Yanjun Han,[1] Zhibiao Hao,[1] Hongtao Li,[1] Yi Luo,[1] Jianchang Yan,[2] Tongbo Wei,[2] Yun Zhang,[2] and Junxi Wang[2]

[1]*Tsinghua National Laboratory for Information Science and Technology, State Key Lab on Integrated Optoelectronics, Department of Electronic Engineering, University, Beijing 100084, China*
[2]*Research and Development Center for Semiconductor Lighting, Institute of Semiconductors, Chinese Academy of Sciences, Beijing 100083, China*

Development of chip-scale optical frequency comb with the coverage from ultra-violet (UV) to mid-infrared (MIR) wavelength is of great significance. To expand the comb spectrum into the challenging UV region, a material platform with high UV transparency is crucial. In this paper, crystalline aluminum nitride (AlN)-on-sapphire film is demonstrated for efficient Kerr frequency comb generation. Near-infrared (NIR) comb with nearly octave-spanning coverage and low parametric threshold is achieved in continuous-wave pumped high-quality-factor AlN microring resonators. The competition between stimulated Raman scattering (SRS) and hyperparametric oscillation is investigated, along with broadband comb generation via Raman-assisted four-wave mixing (FWM). Thanks to its wide bandgap, excellent crystalline quality as well as intrinsic quadratic and cubic susceptibilities, AlN-on-sapphire platform should be appealing for integrated nonlinear optics from MIR to UV region.

## I. INTRODUCTION

Optical frequency comb generation in continuous-wave (c.w.) pumped microresonators via four-wave mixing (FWM) (i.e., Kerr frequency comb) has attracted significant attention, because of its compact size, flexible comb spacing, wide bandwidth, and potential for monolithic integration [1–11]. Since its first demonstration in silica microtoroid [1], a myriad of material platforms have been tested for efficient Kerr comb generation [2–11]. Among them, planar integrated schemes with high robustness and insensitivity to environmental variation are of particular interest, and have been realized with Hydex glass [4], silicon nitride ($Si_3N_4$) [5, 6], aluminum nitride (AlN) [7], diamond [8], silicon [9, 10], and aluminum gallium arsenide (AlGaAs) [11].

Currently, most of Kerr frequency combs are generated in the near-infrared (NIR) region. For a plethora of applications, it is highly desirable to expand the comb coverage into shorter and longer wavelengths, where distinct properties of these material platforms should be considered. For instance, silicon on insulator (SOI) has proved as an excellent platform for mid-infrared (MIR) comb generation [9, 10], owing to its significant cubic susceptibility $\chi^{(3)}$, CMOS compatibility and negligible two-photon absorption (TPA) at MIR regime. On the other hand, visible comb is accessible by converting NIR comb into red and green regions via second-harmonic (SH), sum-frequency (SF) and third-harmonic (TH) processes in sputtered AlN [12] and $Si_3N_4$ [13, 14] microrings. Nevertheless, ultra-violet (UV) comb generation is still elusive, even though frequency conversion of c.w. pumped visible comb could be viable, and the crucial anomalous group velocity dispersion engineering at visible wavelength can be attained with avoided mode crossing [15] or photonic compound configurations [16]. A major obstacle is the relatively large absorption and scattering loss of current material platforms at UV region. Thus, a novel platform with high UV transparency and the combination of inherent $\chi^{(2)}$ and $\chi^{(3)}$ susceptibilities is compelling.

Within the past few years, crystalline AlN has emerged as a promising candidate for UV nanophotonics [17–20], thanks to its enormous direct bandgap (~6.2 eV at room temperature) as well as the mature material growth and processing techniques. Bandgap engineering of III-nitride alloys has enabled UV light emission [21, 22] and detection [23], revealing the potential to realize complex UV photonic circuits with optical sources, routers and detectors integrated on the same chip. Although AlN can be epitaxially grown on different substrates (e.g. sapphire, silicon carbide, and silicon) [24], AlN-on-sapphire wafer features the advantage to naturally formed waveguide structure ($n_{\text{sapphire}}$ = ~1.75 at 1.55 μm) for light manipulation along with the capability of monolithic integration. Compared with polycrystalline AlN on insulator, AlN grown on sapphire exhibits superior crystalline quality, and is less susceptible to scattering and absorption losses at grain boundaries [25, 26], thereby allowing low-loss waveguide components, especially at short wavelengths. As a result, it is expected to alleviate the increased absorption for second harmonic (SH) signal below 400 nm [27] and the high propagation loss in UV region [28] observed in sputtered AlN. Actually, SH generation at UV wavelength has recently been confirmed in AlN-on-sapphire straight waveguides [19] and microring resonators [20]. Additionally, AlN-on-sapphire film exhibits improved physical properties, including high thermal conductivity and robustness [29], making it reliable for high-power handling. Despite the significant advance of crystalline AlN at UV region, its Kerr optical nonlinearity, to our knowledge, has not been explored up to now.

* czsun@tsinghua.edu.cn

In this paper, we demonstrate AlN-on-sapphire as a novel platform for efficient Kerr frequency comb generation, as exemplified at NIR region, by successfully addressing the challenges inherent to nitride processing. High-quality AlN film with the desired thickness for wideband anomalous dispersion engineering is prepared by epitaxial growth technique. An AlN microring with a high loaded quality factor ($Q_L$) of 1.1 million and corresponding cavity buildup of ~432 are attained, yielding a low hyperparametric oscillating threshold of ~25 mW. By taking advantage of dispersive-wave emission, self-referenceable comb spectrum with the coverage of ~1000 nm is achieved. The influence of stimulated Raman scattering (SRS) on hyperparametric oscillation in crystalline AlN is also investigated. Thanks to its intrinsic $\chi^{(2)}$ and $\chi^{(3)}$ susceptibilities as well as excellent optical properties, AlN-on-sapphire is promising for integrated nonlinear optics, especially at elusive UV wavelength.

## II. DEVICE DESIGN AND FABRICATION

### A. Device design

Figure 1(a) presents the top view of our device structure and the principle for Kerr frequency comb generation. A high-power c.w. light is resonantly fed into the microring, which gives rise to degenerate and non-degenerate FWM, transferring the pump power into equidistant spectral lines overlapped with cavity modes [30]. In such a $\chi^{(3)}$ nonlinear parametric process, momentum conservation is intrinsically satisfied, as the involved optical modes are the eigenstates of the microring. On the other hand, energy conservation requires efficient phase matching between the involved modes, which can be fulfilled by employing appropriate anomalous linear dispersion to compensate nonlinear phase shift induced by self- and cross-phase modulation around the strong pump light [8]. Since the bulk material generally exhibits a normal dispersion profile at its transparency window, avoided mode crossing can be utilized to acquire local anomalous dispersion for efficient FWM, albeit with a limited bandwidth [15]. In contrast, expanded anomalous dispersion range for broadband Kerr comb generation is accessible by tailoring the waveguide dispersion to dominate over weak material dispersion at longer wavelength [5].

Figure 1(b) illustrates the group velocity dispersion (GVD) profiles for the AlN microring calculated with finite element method (FEM). Here, GVD is given by $-(\lambda/c)\cdot(d^2 n_{\text{eff}}/d\lambda^2)$, with $n_{\text{eff}}$, $\lambda$, and $c$ being the effective index, wavelength and light speed in vacuum, respectively. In our simulation, the material dispersion of bulk AlN is determined by Sellmeier formula based on ellipsometer measurement, while the waveguide dispersion takes the waveguide geometry, such as microring cross section, bending radius and sidewall slope angle, into consideration. It is found that a wideband anomalous dispersion range (GVD > 0) from NIR to MIR region is secured for fundamental transverse-magnetic ($TM_{00}$) mode by tailoring AlN microring height (H) to 1.2 μm, while keeping its width and outer radius at at 3.5 and 60 μm, respectively. Such device geometry also has the potential for MIR Kerr comb generation from 2 to 3 μm, benefiting from the extended anomalous GVD range and the transparency of AlN in MIR region [31].

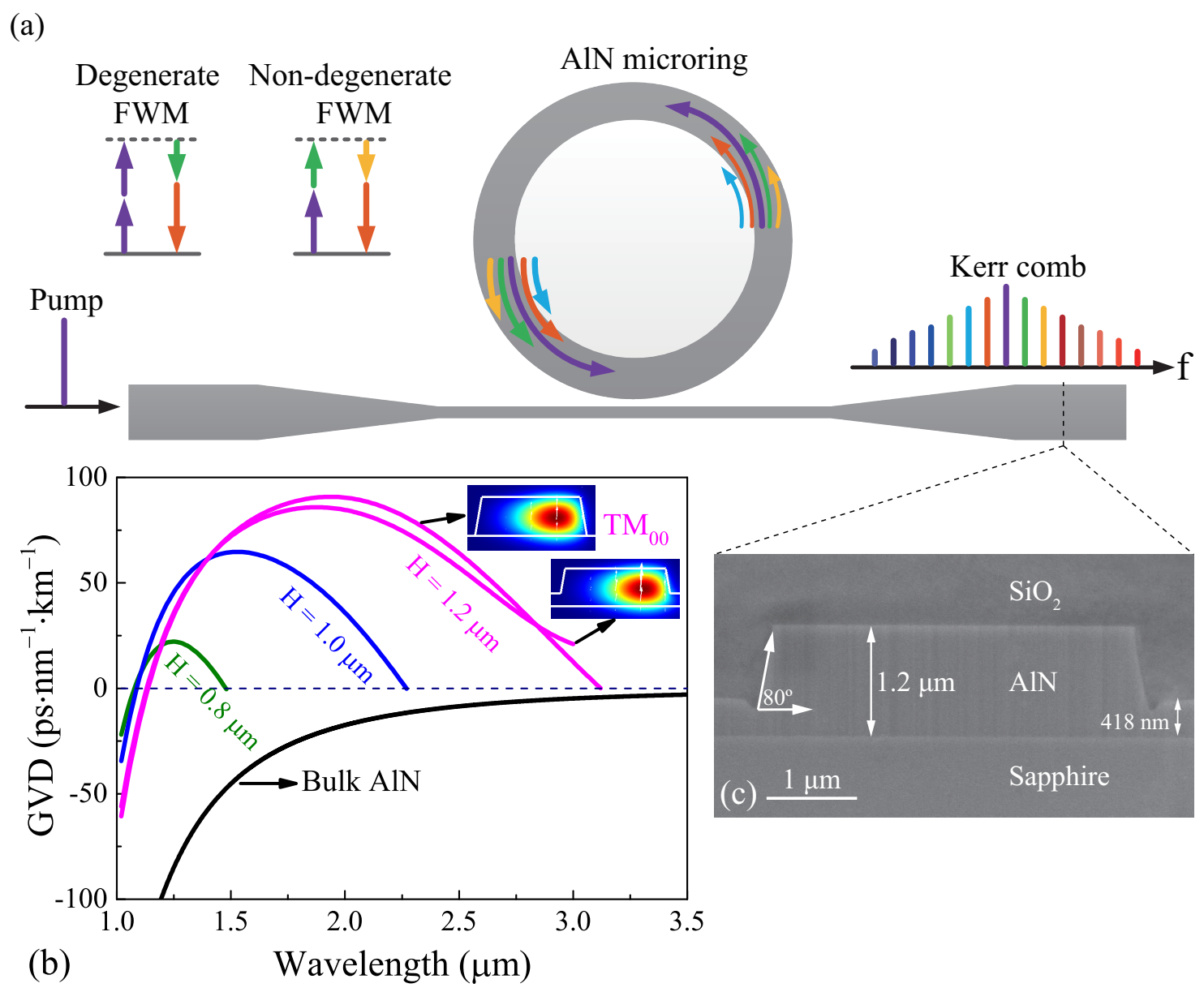

FIG. 1. (a) Schematic illustration of the adopted AlN-on-sapphire microring structure and the principle for Kerr comb generation. (b) Calculated group velocity dispersion (GVD) of bulk AlN as well as that of $TM_{00}$ mode in the AlN microring. The radius and width of the microring are kept at 60 and 3.5 μm, while the height varies from 0.8 to 1.2 μm. Insets: simulated mode profiles at 1.55 μm for completely and partially etched microrings, respectively. (c) Scanning electron microscopy (SEM) image of the cleaved chip end facet with 80° sidewall slope angle resulting from the dry etching process.

The relatively low etching rate of AlN, as a result of its high bond energy [32], makes it challenging to process such a thick AlN film for exact nanopattern transfer. Fortunately, it is noted that the GVD profile of a partially etched microring deviates only slightly from that of a completely etched one [see the magenta curves of Fig. 1(b)], thereby allowing the adoption of a partially etched waveguide to facilitate device fabrication. To ensure a wide anomalous GVD profile, a relatively large width of 3.5 μm is adopted for the microring, which would give rise to multiple transverse mode families. To ensure the coupling ideality, a single-mode bus waveguide with an optimized width of 1.35 μm is selected for phase matching with fundamental modes inside the microring. This facilitates injection of c.w. pump light into the microring and effective extraction of comb lines from the resonator via evanescent field coupling.

### B. Device fabrication

In the experiment, single-crystalline AlN film with a thickness of 1.2 μm is prepared on *c*-plane (0001) sapphire substrate by metal organic chemical vapor deposition (MOCVD) [33], along with the measured refractive index of ~2.1 at 1.55 μm. After depositing $SiN_x$ mask on the wafer, the microring and associated bus waveguide are defined by electron beam lithography (EBL) with ZEP520 resist, followed by pattern transfer with reactive ion etching (RIE) and optimized $Cl_2/BCl_3$/Ar-based inductively coupled plasma (ICP) etching [32]. Finally the device is embedded with a 3-μm-thick silicon dioxide ($SiO_2$) by plasma enhanced chemical vapor deposition (PECVD), and no additional annealing treatment is performed.

Figure 1(c) shows the cross section of the cleaved chip end facet. Here, the bus waveguide is laterally tapered to a width of 4 μm to ensure a better modal overlap with our lensed fibers (3.5 μm mode field diameter) for enhanced fiber-to-chip coupling efficiency. Partially etched waveguide with a 418-nm-thick unetched AlN layer at the ridge bottom is identified, which relaxes the required etching selectivity for exact EBL pattern transfer into thick AlN film. Meanwhile, the enhanced waveguide-to-microring coupling allows a coupling gap as wide as 800 nm, thereby greatly alleviating the fabrication tolerance [34]. Thanks to its high thermal conductivity (320 $W\cdot m^{-1}K^{-1}$), improved heat dissipation can also be implemented through the unetched AlN layer, which facilitates thermal connection of the microring to its surrounding region.

## III. KERR FREQUENCY COMB FORMATION

### A. Hyperparametric oscillation

The transmission spectrum shows that the fiber-to-chip insertion loss of our device is as low as ~3 dB/facet. Two sets of TM mode families are excited inside the microring and the free spectral range (FSR) for $TM_{00}$ mode is 369 GHz at telecom band. According to Fig. 2(a), in which a high-resolution $Q$ factor measurement is depicted, the recorded full-width at half-maximum (FWHM) linewidth ($\Delta f_{\mathrm{FWHM}}$) around 1558.1 nm for $TM_{00}$ mode is ~176 MHz, revealing a $Q_L$ of 1.1 million. The intrinsic and coupling $Q$ factors ($Q_{\mathrm{int}}$ and $Q_C$) are extracted accordingly, as the microring is under-coupled. The cavity buildup is thereby estimated to be ~432 based on the formula: $P_{\mathrm{cir}}/P_{\mathrm{in}} = \mathrm{FSR}/(\pi\Delta f_{\mathrm{FWHM}})\cdot 2Q_L/Q_C$. Figure 2(b) illustrates the calculated wavelength-dependent $Q_C$ for our device [34], suggesting a good agreement with experimentally extracted value in Fig. 2(a). The insets present the simulated mode profiles of the microring. It is evident that the microring exhibits a tighter optical confinement at shorter wavelengths, which hampers the waveguide-to-microring coupling and can lead to a deteriorated Kerr comb extraction efficiency.

Figure 2(c) illustrates a typical initial state of Kerr comb generation, captured by an optical spectrum analyzer (OSA). Pump light with $P_{\mathrm{in}}$ ~31 mW is gradually tuned into resonance near 1558.1 nm and self-stabilized by thermal locking [35], which triggers the onset of hyperparametric oscillation with the first ($1^{\mathrm{st}}$) sidebands 9 FSRs away from the pump line. The formation of such multiple-FSR spaced comb spectrum is determined by the dispersion and resonance linewidth of the microring [36]. By plotting $1^{\mathrm{st}}$ mode power as a function of $P_{\mathrm{in}}$ [see Fig. 2(d)], a low hyperparametric threshold $P_{\mathrm{th}}$ of ~25 mW is identified. Upon further enhancing $P_{\mathrm{in}}$, the $1^{\mathrm{st}}$ mode power become clamped due to pump power transfer to newly-generated sidebands through cascaded FWM.

Based on the recorded threshold power, Kerr nonlinear coefficient $n_2$ of AlN-on-sapphire film is determined by [8]:

$$P_{\mathrm{th}} \approx 1.54\frac{\pi}{2}\cdot\left(\frac{Q_C}{2Q_L}\right)\cdot\frac{n_0^2\cdot V_{\mathrm{eff}}}{n_2\cdot\lambda_0\cdot Q_L^2} \tag{1}$$

where $\lambda_0$ is the pump wavelength, $n_0$ is the refractive index of AlN, and $V_{\mathrm{eff}}$ is the effective mode volume of the microring calculated with FEM. To ensure the validity of our estimation, thresholds of different devices with gap size of 0.7 and 0.75 μm are also included. Taking into account the measured $Q$ factors, $n_2$ of crystalline AlN at telecom band is deduced to be ~(3.5 ± 0.8) × $10^{-19}$ $m^2 \cdot W^{-1}$, which is on a par with the value reported for sputtered AlN [7].

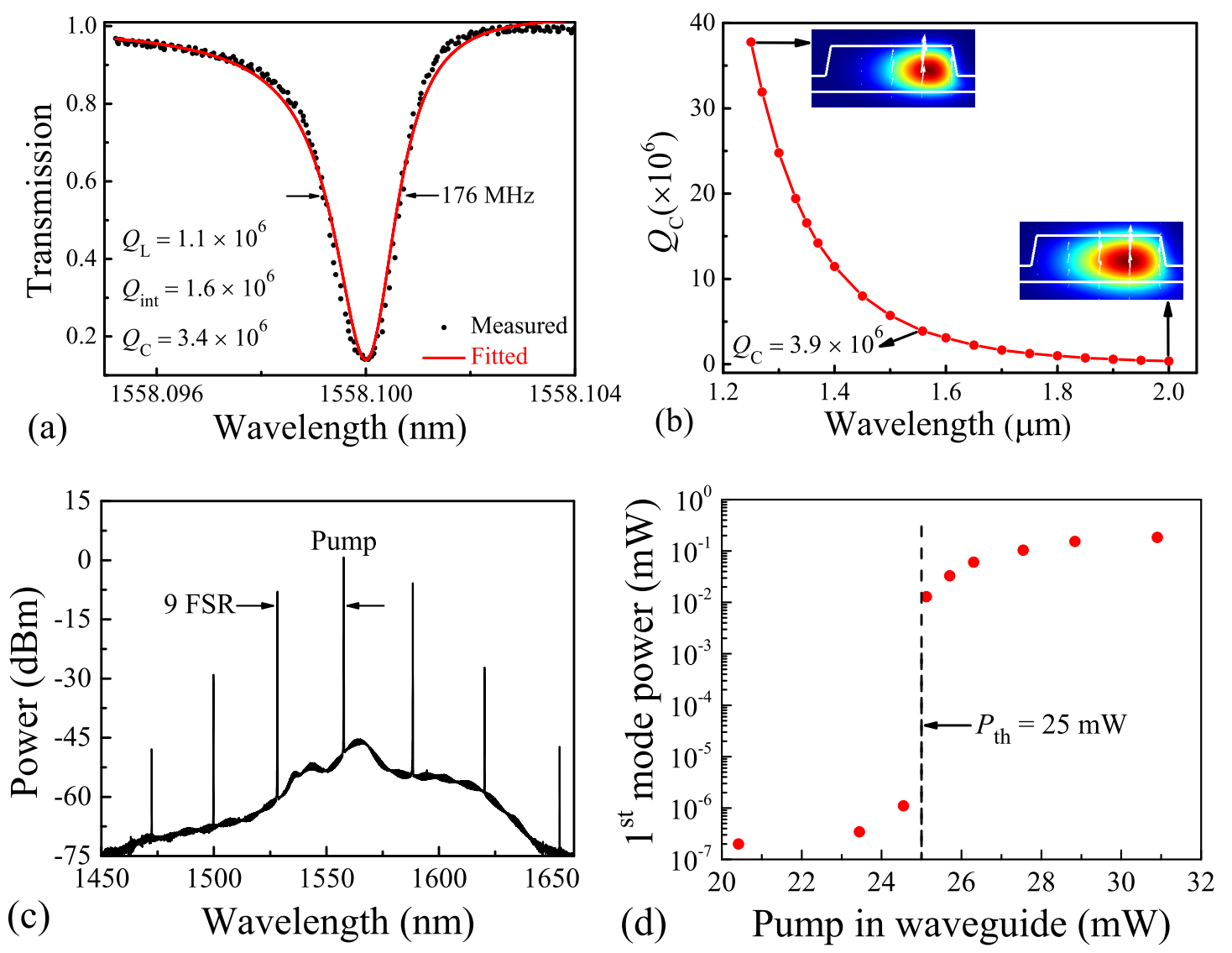

FIG. 2. (a) Resonance linewidth of $TM_{00}$ mode at 1558.1 nm along with the extracted $Q$ factors. (b) Calculated $Q_C$ versus wavelength for $TM_{00}$ mode of the microring, indicating a reduced coupling strength at shorter wavelengths. Inserts: simulated mode profiles of the microring at 1.25 and 2 μm, respectively. (c) The formation of primary comb lines with 9 FSRs spacing at a $P_{in}$ ~31 mW. (d) A log plot of 1st oscillating mode power versus $P_{in}$, exhibiting a hyperparametric threshold of 25 mW.

## B. Broadband Kerr comb generation

By further enhancing $P_{in}$, secondary comb lines emerge around the primary sidebands via higher-order degenerate and non-degenerate FWM processes [36]. With a $P_{in}$ of 1 W, which is limited by the available power of our erbium doped fiber amplifier (EDFA), a nearly octave-spanning comb spectrum covering from ∼1075 to 2075 nm with a single FSR spacing of 369 GHz is achieved, as illustrated in Fig. 3. The output spectrum is recorded by two OSAs with the spectral range of 0.6 – 1.75 μm and 1.2 – 2.4 μm, respectively. It is inferred that the generated AlN comb spectrum should be sufficiently wide for frequency stabilization via 2f-3f self-referencing [6]. The 3-dB bandwidth ($\Delta f_{3dB}$) of Kerr frequency comb reads [37]:

$$\Delta f_{3dB} = \frac{0.315}{1.763}\sqrt{\frac{2\gamma P_{in}\mathcal{F}}{\pi|\beta_2|}} \tag{2}$$

where $\gamma = 2\pi n_2/(\lambda A_{eff})$ is the nonlinear parameter ($A_{eff}$ being the effective mode area of the microring), $\mathcal{F} = f_0/\Delta f_{FWHM}$ is the cavity finesse ($f_0$ being the pump frequency), and $\beta_2$ is the second-order (2nd) GVD parameter. For our AlN microring, $\gamma$ = 676 $W^{-1} \cdot km^{-1}$, $\mathcal{F}$ = 2096, $\beta_2$ = −98.7 $ps^2 \cdot km^{-1}$. Thus, $\Delta f_{3dB}$ is deduced to be 17 THz, which is in agreement with the extracted value (∼13 THz) from recorded comb spectrum in Fig. 3.

Meanwhile, in contrast to the red-shifted dispersive-wave emission reported in Ref. [6], blue-shifted one is observed around 1130 nm in Fig. 3 due to the presence of normal third-order dispersion (TOD), which helps extend comb spectrum into normal GVD region at shorter wavelengths. Actually, the comb spectrum coverage within the microring can be even wider, when taking into account the non-optimized fiber-to-chip coupling away from telecom wavelength and the reduced straight waveguide-to-microring coupling at shorter wavelengths [see Fig. 2(b)]. By filtering a portion of comb spectrum for radio-frequency (RF) measurement, a broad RF beat note is identified at low frequencies, indicating a typical high-noise state of the generated Kerr comb [36]. For practical applications, formation of temporal dissipative Kerr soliton with high coherence is preferred, which can be accessible by utilizing the 'pump kicking' technique [6], and will be the topic of our following work.

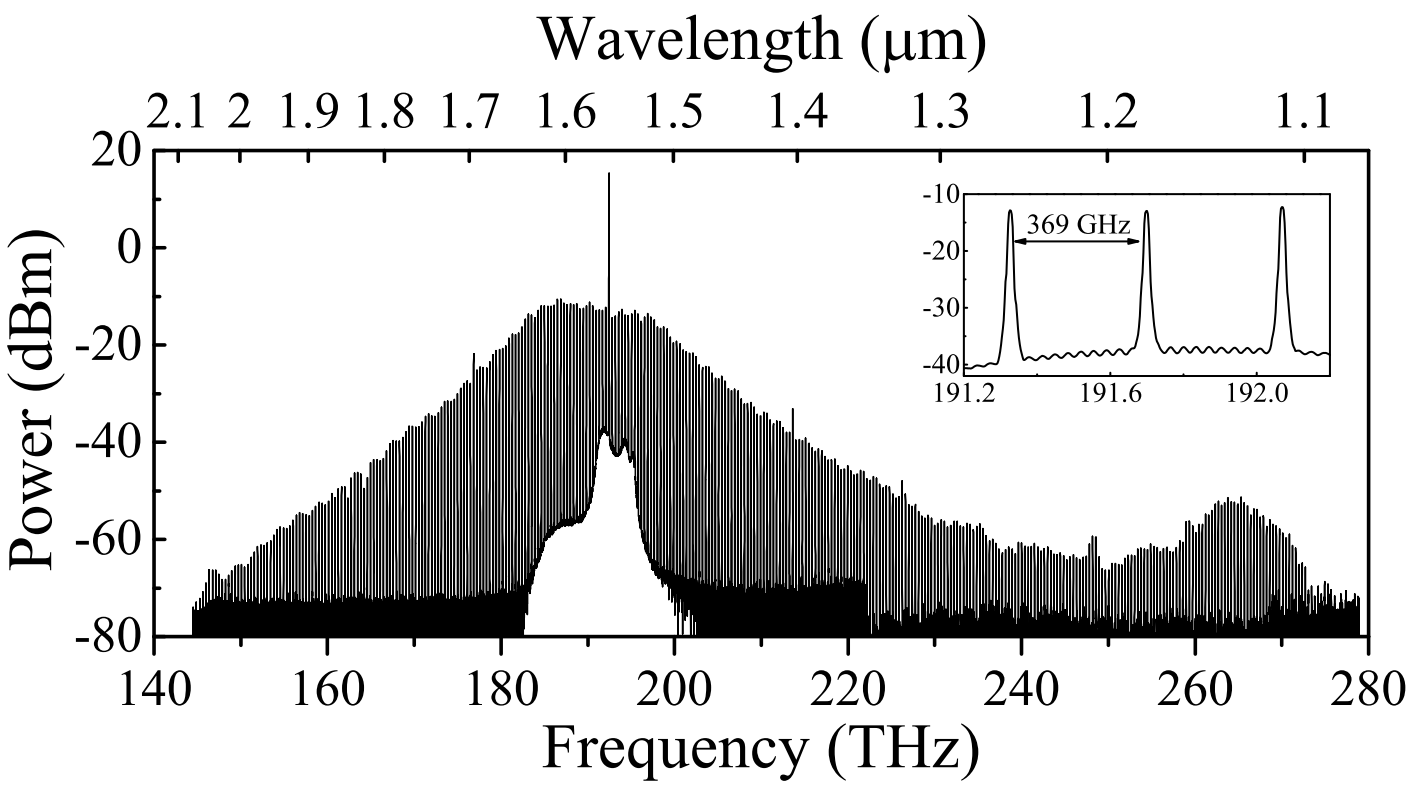

FIG. 3. Recorded nearly octave-spanning Kerr comb spectrum spanning from ~1075 to 2075 nm at $P_{in}$ of 1 W, along with the observation of blue-shifted dispersive-wave emission near 1130 nm. Insert: zoom-in comb lines, revealing a single FSR spacing of ~369 GHz.

## IV. THE INFLUENCE OF RAMAN EFFECT ON KERR FREQUENCY COMB GENERATION

Crystalline AlN-on-sapphire film normally exhibits narrow-linewidth phonon spectra and would give rise to SRS [38], competing with Kerr nonlinearity-induced hyperparametric oscillation. Of these two distinct $\chi^{(3)}$ nonlinear optical processes, microcavity-based hyperparametric oscillation originates from the quasi-instantaneous response of electronic clouds to optical field and requires appropriate anomalous GVD for phase matching. In contrast, phase matching is intrinsically satisfied in SRS, as it results from delayed response of pump light to lattice vibration and the involved optic phonons are excited around the centre of Brillouin zone. Since higher-order modes of the microring typically exhibit a larger GVD value than that of fundamental modes, it can be utilized to investigate the competition between SRS and Kerr nonlinearity in AlN.

In our experiment, microrings with a radius of 100 μm and associated gap size of 600 nm are fabricated, featuring high $Q$ factors for both fundamental ($TM_{00}$) and first order ($TM_{10}$) modes, corresponding to cavity buildup of 303 and 201, respectively [see Fig. S1 in supplementary material]. Figure 4(a) illustrates the calculated GVD profiles for both $TM_{00}$ and $TM_{10}$ modes, where wideband anomalous GVD is achieved at telecom band, along with a notably larger GVD value for $TM_{10}$ mode. Figure 4(b) depicts the Raman lasing spectrum by selective excitation of $TM_{10}$ mode with TM-polarized pump around 1552.634 nm. With $P_{in}$ ~1 W, Stokes emission with a Raman shift of 612 $cm^{-1}$ away from the pump is recorded, implying the excitation of $A_1^{TO}$ phonon in AlN [38]. On the other hand, broadband Kerr comb generation is achieved by pumping $TM_{00}$ mode of the microring around 1551.9 nm [see Fig, 4(c)], and no Raman lines show up as the pump is tuned into resonance. Our results indicate that the mode dispersion makes a big difference for the occurrence of SRS or hyperparametric oscillation in AlN-on-sapphire microrings, which is in accordance with tailoring cavity geometry for effective FWM gain in silica microcavity [39].

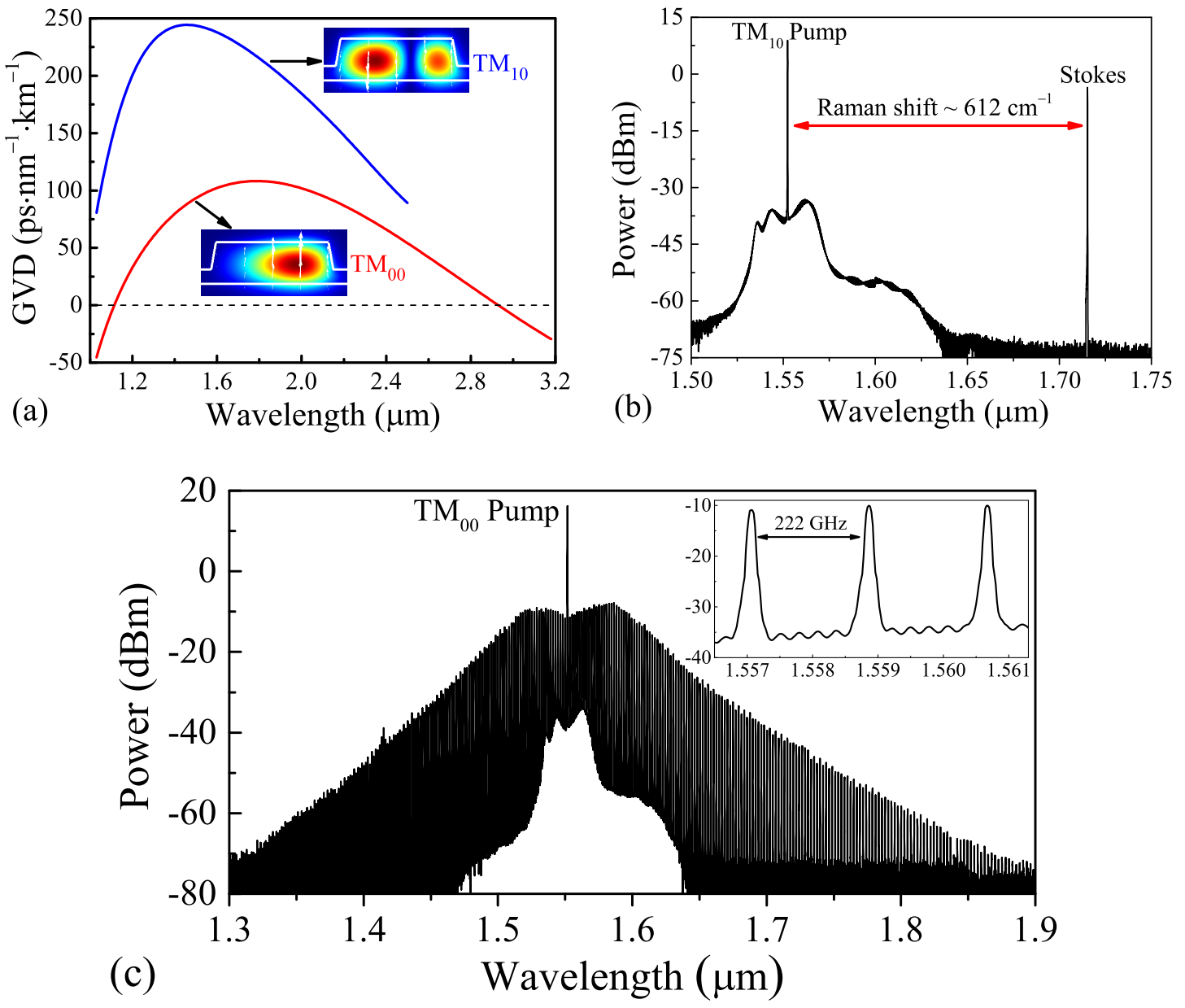

FIG. 4. (a) Calculated GVD profiles and simulated mode fields at 1.55 μm (insets) for $TM_{00}$ and $TM_{10}$ modes of the microring (radius: 100 μm), respectively. (b) Recorded Stokes lasing spectrum with a Raman shift of 612 $cm^{-1}$ from the pump frequency when pumping $TM_{10}$ mode at $P_{in}$ ~1 W. (c) Kerr comb formation with a single FSR spacing of 222 GHz when pumping $TM_{00}$ mode around 1551.9 nm ($P_{in}$ ~1 W).

Apart from the influence of cavity dispersion, it is found that transition from SRS to FWM can also be achieved by adjusting intracavity power of AlN microrings endowed with appropriate dispersion profile. In this case, a device with a radius of 80 μm and a gap size of 700 nm is employed. Figure 5(a) depicts the simulated GVD profiles for fundamental $TM_{00}$ and $TE_{00}$ modes, respectively, where $TE_{00}$ mode exhibits a much smaller GVD value. In our previous work, highly efficient 1st and 2nd Stokes lasing are confirmed for this device with $TM_{00}$ pump, and FWM is absent even at a high pump level [38]. Here, $TE_{00}$ mode with a FSR of ~286 GHz at telecom band is selected. The measured $\Delta f_{FWHM}$ is ~211 MHz for the resonance at 1559.659 nm, indicating a $Q_L$ of ~0.9 million and a cavity buildup of ~353 [see Fig. S2 in supplementary material]. As TE-polarized pump is tuned into resonance with $P_{in}$ ~126 mW, Raman lasing occurs with typical 1st and 2nd Stokes spectrum recorded in Fig. 5(b), while FWM is absent. The Raman shift is found to be ~660 $cm^{-1}$ (19.8 THz), revealing the excitation of $E_2^{high}$ phonon in AlN [38].

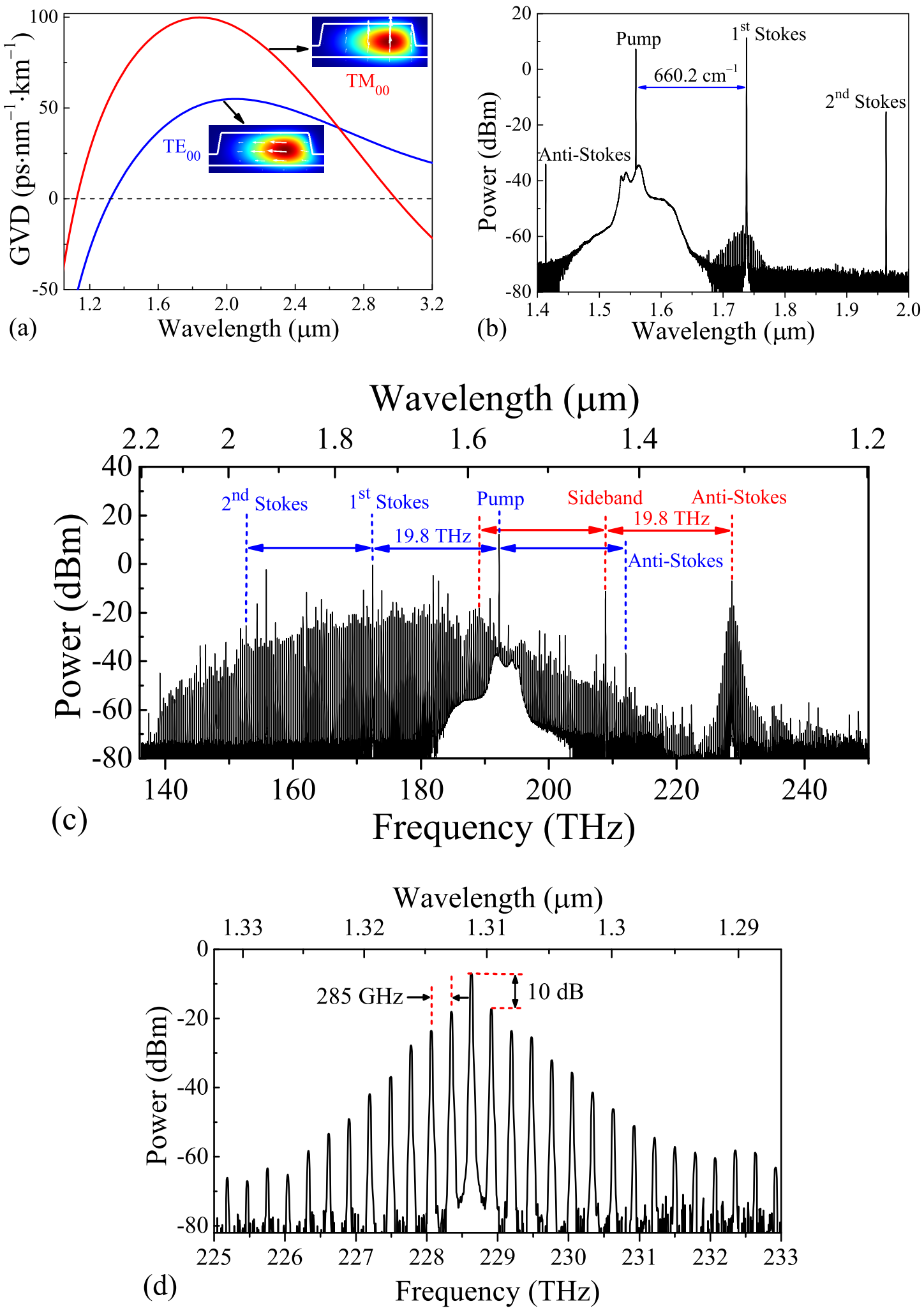

FIG. 5. (a) Calculated GVD profiles of $TM_{00}$ and $TE_{00}$ modes inside the microring (radius: 80 μm). Insets: mode fields of the microring at 1.55 μm. (b) Cascaded Raman lasing spectrum with the Raman shift of ~660 $cm^{-1}$ away from the pump frequency for $TE_{00}$ pump and $P_{in}$ of ~126 mW. (c) Recorded broadband frequency comb spectrum with the coverage from 1.2 to 2.2 μm for enhanced $P_{in}$ of ~1 W. Typical Raman lines are identified with the separation of 19.8 THz. (d) A zoom-in spectrum of the strong peak around 1311 nm, exhibiting a single FSR spacing of 285 GHz and 10-dB power contrast between the peak center and its first sideband.

By boosting $P_{in}$ to ~1 W, SRS occurs when the pump detuning is large (i.e., small $P_{cir}$ within the microring). As the pump is gradually tuned into resonance (i.e., enhanced $P_{cir}$), satellite sidebands arise around the pump, Stokes and anti-Stokes lines due to Raman-assisted FWM process [40]. When the pump wavelength finally approaches the cavity resonance, a broadband comb spectrum spanning from 1.2 to 2.2 μm with a single FSR spacing is recorded, as illustrated in Fig. 5(c). The evolution of this comb is presented in Fig. S3 of supplementary material. In contrast to Fig. 3 and Fig. 4(c), the observed comb spectrum in Fig. 5(c) exhibits an asymmetric envelope with respect to the pump line and extends into longer wavelength side. Strong comb lines protruding out of the envelope with a spacing of 19.8 THz are identified, which is in accordance with the recorded Raman shift in Fig. 5(b), indicating the existence of SRS process. The underlying mechanism may be ascribed to the relatively small GVD value of $TE_{00}$ mode, which can be compensated by the self- and cross-phase modulation inside the microring resulting from strong intracavity power, thus enabling effective FWM gain. This is similar to the prior demonstration in silica microtoriod by altering the cavity loading (i.e., intracavity power) to control the transition between FWM and Raman oscillations [41].

Additionally, it is noteworthy that a strong peak with narrow spectral range emerges around 1311 nm, helping extend the comb spectrum into shorter wavelengths [see Fig. 5(c)]. Interestingly, the separation between the peak center and a strong FWM sideband (indicated by red label) is equal to the Raman shift (~19.8 THz). This coincidence suggests the peak formation might be due to coherent anti-Stokes Raman scattering (CARS) [42, 43], which transfers the power of a strong FWM sideband and corresponding Stokes line to anti-Stokes line (indicated by red label). This is supported by the 10-dB power contrast between the peak center and its first sideband [see Fig. 5(d)], implying a narrow-band amplification process, as determined by the phase matching of involved modes as well as the narrow Raman linewidth of $E_2^{high}$ phonon in AlN (~114 GHz) [38]. Meanwhile, it is noted that the formation of the symmetric envelope around 1311 nm is sensitive to the detuning of pump wavelength, indicating a phase matching-dependent process [43]. In our future work, further investigation will be performed for a definite conclusion.

## V. CONCLUSION AND OUTLOOK

Crystalline AlN has been demonstrated for broadband NIR Kerr frequency comb generation. Nearly octave-spanning comb spectrum is recorded in high-$Q$ AlN-on-sapphire microrings, as a consequence of the tailored cavity dispersion as well as the occurrence of blue-shifted dispersive-wave emission. The influence of SRS on Kerr comb formation is investigated by altering the dispersion profile or intracavity power. Through a Raman-assisted FWM process, broadband comb spectrum is achieved along with the presence of strong Raman lines. Our work paves the way to exploiting Kerr optical nonlinearity in AlN-on-sapphire film for efficient on-chip frequency comb generation. Thanks to the noncentrosymmetric structure of AlN, high-speed electrical switching frequency comb is anticipated with electro-optic Pockels effects [44]. Likewise, visible comb is accessible by converting NIR comb into visible region via SH and SF process [12]. Furthermore, its large bandgap and excellent crystalline quality permit c.w. pumped visible comb generation, which provides an access for UV comb formation via frequency conversion schemes. It is inferred that AlN-on-sapphire film should have immense potential for the applications in integrated nonlinear optics.

## FUNDING INFORMATION

This work was supported in part by National Basic Research Program of China (2014CB340002); National Natural Science Foundation of China (61210014, 61621064, 61307024, 61574082, 51561165012); High Technology Research and Development Program of China (2015AA017101); Tsinghua University Initiative Scientific Research Program (20131089364, 20161080068, 20161080062); and Open Fund of State Key Laboratory on Integrated Optoelectronics (IOSKL2014KF09).

## ACKNOWLEDGMENTS

The authors would like to thank Prof. Changxi Yang of Tsinghua University for his help in optical spectrum measurements.

---

# Supplementary Material: Broadband frequency comb generation in aluminum nitride-on-sapphire microresonators

Xianwen Liu,[1] Changzheng Sun,[1] Bing Xiong,[1] Lai Wang,[1] Jian Wang,[1] Yanjun Han,[1] Zhibiao Hao,[1] Hongtao Li,[1] Yi Luo,[1] Jianchang Yan,[2] Tongbo Wei,[2] Yun Zhang,[2] and Junxi Wang[2]

[1]*Tsinghua National Laboratory for Information Science and Technology, State Key Lab on Integrated Optoelectronics, Department of Electronic Engineering, University, Beijing 100084, China*

[2]*Research and Development Center for Semiconductor Lighting, Institute of Semiconductors, Chinese Academy of Sciences, Beijing 100083, China*

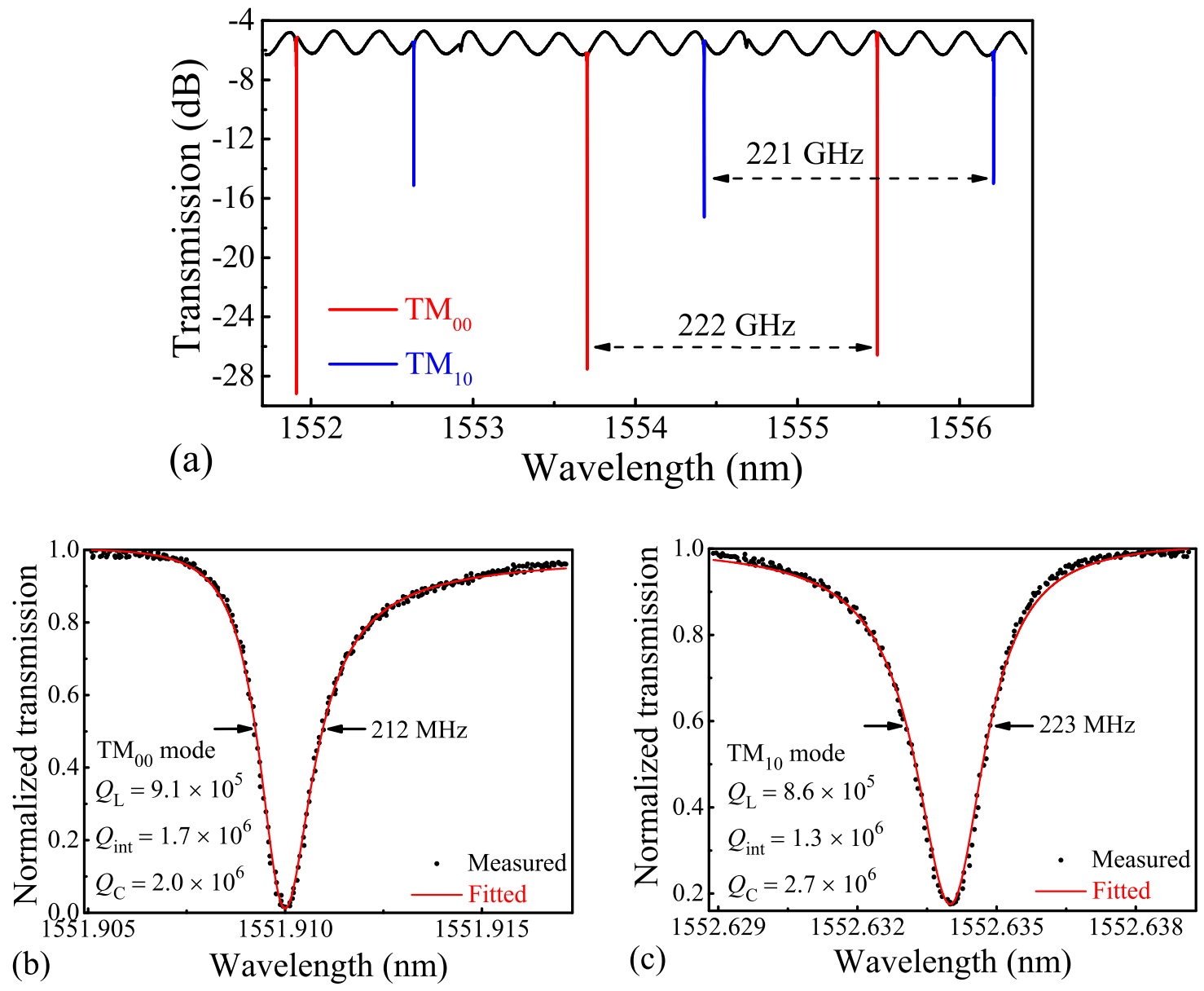

FIG. S1. Characterization of the microring (radius: 100 μm, gap: 600 nm). (a) Transmission spectrum with a low insertion loss and large on-resonance extinction ratio for both $TM_{00}$ and $TM_{10}$ modes, respectively. (b) and (c) Measured resonance linewidth and extracted $Q$ factors for $TM_{00}$ and $TM_{10}$ modes around 1551.91 and 1552.634 nm, respectively.

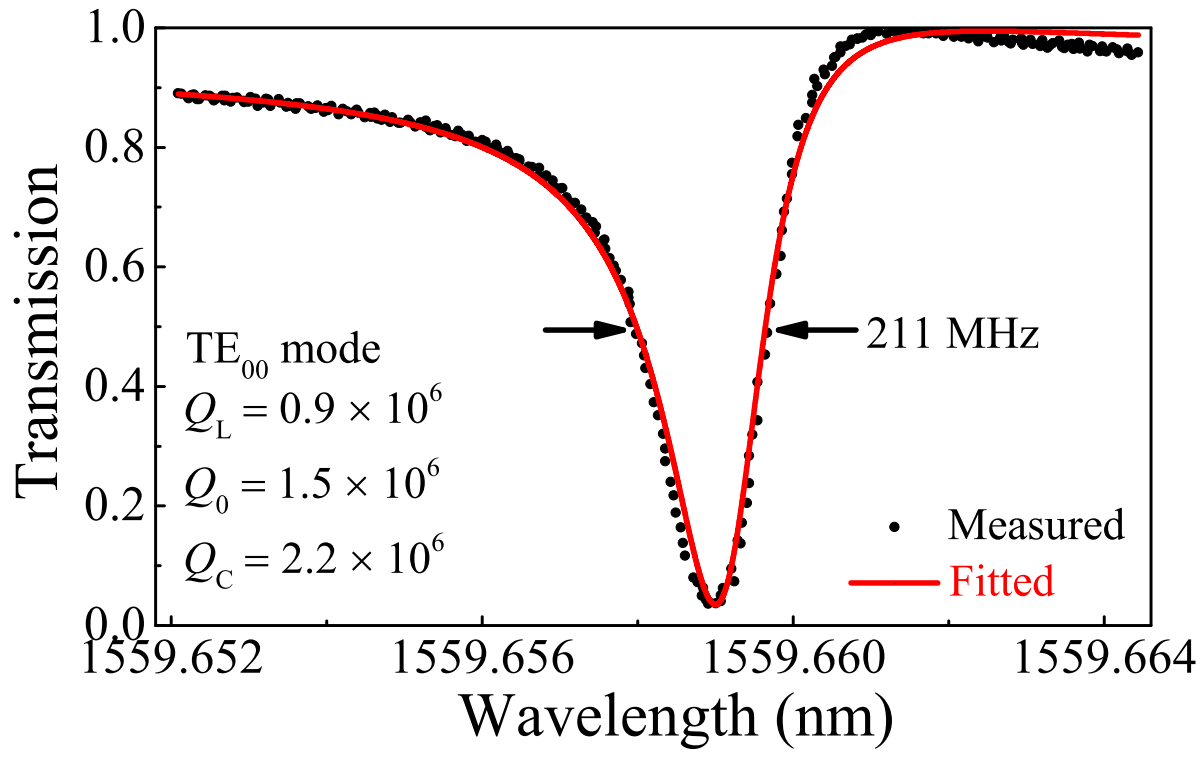

FIG. S2. Measured resonance linewidth and extracted $Q$ factors around 1559.659 nm for $TE_{00}$ mode of the microring (radius: 80 μm, gap: 700 nm).

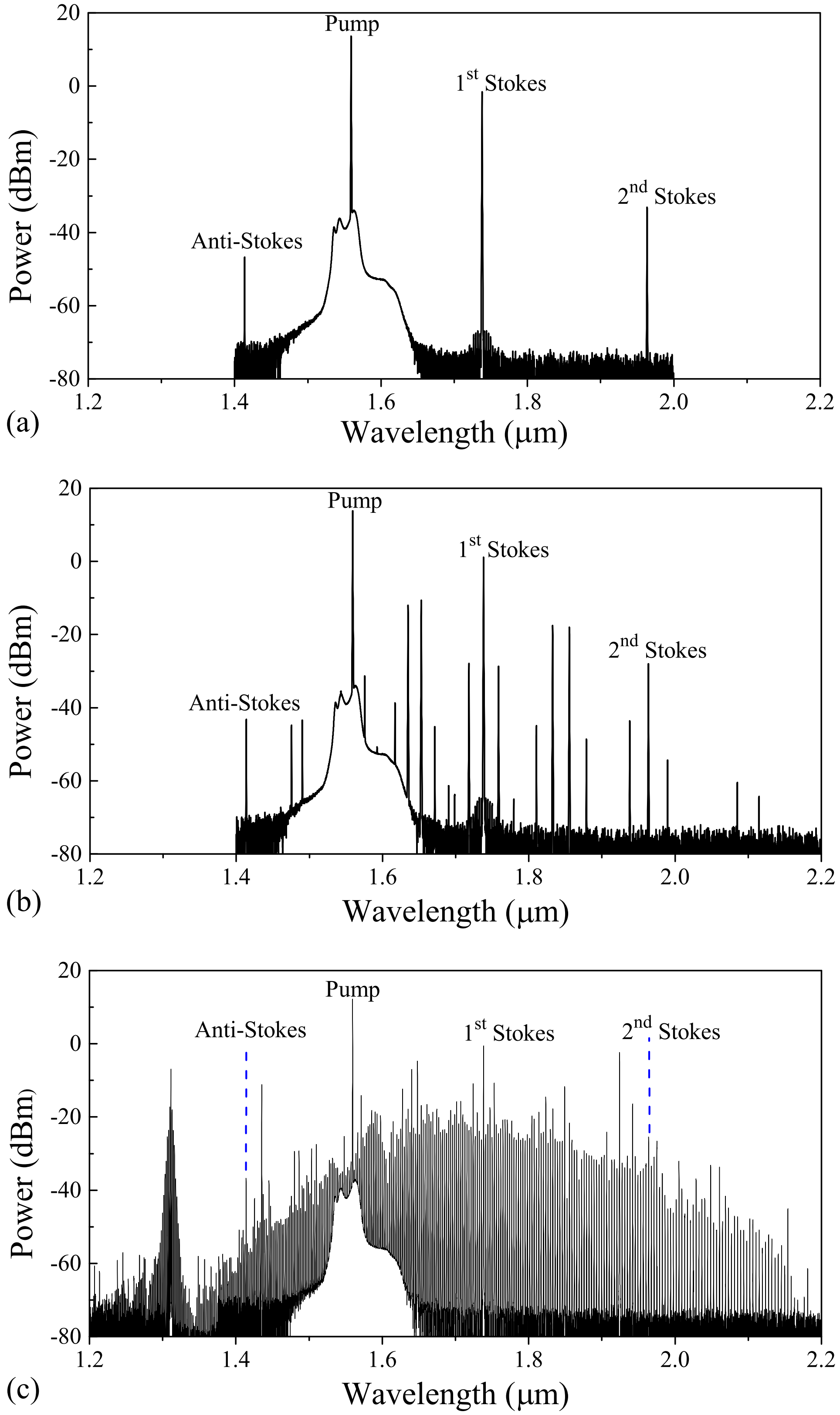

FIG. S3. Spectrum evolution with TE-polarized pump and ∼1 W power in bus waveguide (microring with outer radius of 80 μm). (a) SRS occurs with 1st and 2nd Stokes lines at a large pump detuning. (b) Comb lines formation via Raman-assisted four-wave mixing with further reduced detuning. (c) Broadband comb generation spanning from 1.2 to 2.2 μm with a single FSR spacing when pump wavelength approaches cavity resonance. A strong peak is observed around anti-Stokes frequency, which facilitates the coverage into short wavelength.